# Statistical Approaches for Modelling Cancer Bioassays

Christos P. Kitsos, Nikolaos K. Tavoularis[1], Thomas L. Toulias, George Lolas

Technological Educational Institute of Athens
Department of Mathematics

**Abstract**

This paper discusses the possible ways to analyse the data, adopting a matrix notation, so often used in Bioassays. The paper also reviews the Multistage Models (MM). The MM class of models is applied for extrapolation, to the region of Low-Dose. The effect of covariates in experimental carcinogenesis is introduced and the relative efficiency is evaluated. Certainly the discussed case was refereed to uncorrelated covariates and therefore an open problem might be the multicollinear predictive covariates.Various nonlinear models are discussed, giving more emphasis on the Michaelis-Menten and the Fisher's information for them is discussed.

**Keywords:** Multistage Models, Covariates, Michalis-Menten, Non-linear models, Fisher's Information.

## *1. Introduction*

This paper reviews the Multistage Models (MM) and provides some nonlinear models adopting to Cancer Bioassays. The MM class of models is applied for extrapolation, to the region of Low-Dose, Kitsos (1997) while the general prediction problem has been discussed under a different approach by Kitsos (1993), with some applications to carcinogenesis. The effect of covariate omissions in experimental carcinogenesis is introduced and the relative efficiency is evaluated. In Appendix the Michalis-Menten model and some extensions on it are discussed. A number of graphs of various curves adopting in bioassays are presented in Appendix B, while the Fisher's Information for some models is evaluated in Section 5.

## *2. Descriptive Summary Tables and Analysis*

A common way of presenting summary data, in all the of statistical problems, and therefore in Cancer Epidemiological studies, is through the use of a data matrix, to create a Table, to group the data, in Kendal's suggestion. However, a formal construction of a Table, must be defined in such a way so that to model the logical design underlying of the appropriate statistical data set of a epidemiological study or Bioassay. A summary statistic can be eventually viewed as a mathematical function. The so called "dependent" variable of the function is a numerical variable, which is referred to as the summary variable (elsewhere called summary attribute, and analysis variable, or response variable). A summary variable is defined by a name and a type (e.g., real, integer, nonnegative real, nonnegative integer). The "independent" variables are nominal or range variables, which are referred to as category attributes (elsewhere called categorical variables and classification variables, or input variables).

---

[1] *Died in a car accident on March 17, 2009.*

In principle a category attribute in Biological studies and especial in Cancer Bioassays is defined by a name and a domain. Usually the domain consists from a few values (usually called "codes"), that is the ordinal number of the domain is a small positive integer number.

For example, the categorical variable sex is a two valued category attribute, which is often used to all epidemiological studies, and "acts as covariate" to some of these. Same treatment needs the categorical variable Cancer ("yes" or "no"), while the value "yes" needs further investigation (through another categorical value) concerning the "type" of Cancer.

So if we consider the set $C=\{C_1,.....,C_r\}$ of all category attributes in a given population $\Omega$ we can consider that there is assigning a mapping (or set function) $\mu_j$ from the domain of $C_j$ to the power set of $\Omega$, $j=1,.....,r$. Given a category value $c_j$, then the values of $\mu_j(c_j)$ denotes the set of units of observation matching the condition $C_j = c_j$, and is called the category associated with the category value $c_j$.

Two main assumptions has to be considered, even though there are not stated in Cancer Bioassays problems:

- *Assumption 1*. Partitioning attributes.

Each unit of observation in $\Omega$ falls into at most one elementary category.

- *Assumption 2*. Data additivity.

The summary variable $X$ provides additive total information.

Assumption 1 simply declares that elementary categories are mutually disjoint and cover Ω. In other words, the set of category attributes **C** acts as a classification scheme for the units of observation in Ω. This assumption is usually testing by a $n \times m$ contingency table and the appropriate $X^2$ test. The 2x2 is the most well known and is sometimes linked with the logit model, Kitsos (2007). Assumption 2 declares that the total information for the variable X is the sum of the information provided by each observation of *X*. In principle, by "information" the defined by Fisher information is considered. The augmentation of the data provides "more information" additively, in the sense that the total information is the sum of the information added by the n observations. This certainly needs the independence assumption which is always there, somehow considered by the experimentalist, although not always true. But even is such cases the likelihood function can be evaluated for all the observations, by "pretending" that Ford and Silvey (1980), are independed, Kitsos (1992).

It is widely accepted that the first stage statistical analysis, especially to Biological data analysis is based on a compact "table form" of the data.

Considering a statistical analysis, which contains multiple summary tables over the same population, the following definition is given.

*Definition 2.1*. Two or more summary tables $T_1,...,T_s$ are homogeneous, if they contain data on the same summary variable X for the same population Ω, but use different classification schemes .

The collection of the $T_i$'s is referred to as a polyptych (diptych for $s=3$, triptych for $s=3$) of summary tables on X in Ω.

A problem raised by the management/manipulation of a statistical analysis for the collected data, is when containing a polyptych of summary tables. In such a case the data integration, which consists in viewing the tables of a polyptych as "projections" of a higher-dimensional summary table, called the universal table. As an example it is rather difficult the EU data set from Cancer projected to country XX to provide the exact summary statistics for this particular country.

For a universal-scheme interface to be practical, the polyptych under examination has to be consistent. That means that there exist universal tables, that is, a summary table with classification scheme **U**, whose "projection" on the **R**i's return the tables of the polyptych. This needs a certain organization of the data sets, which is not so happen in practice, although various research centers on Cancer attempts so. In order to state the notions of a universal table and consistency precisely, we introduce two notions: that of a marginal of a summary table and that of a universal category relation.

*Definition 2.2.* A universal category relation is a category relation over the universal classification scheme **U** such that its projections onto the $\mathbf{R}_i$'s restore the category relations $\mathbf{R}_i$'s of the tables of the polyptych.

An interesting case occur whenever two or more category attributes refer to a common, but not identical categorization criterion, and therefore is more or less tightly connected. This is a weak point of the statistical analysis of the Bioassay. That is we would recommend a pilot study before any statistical data analysis. Based on this knowledge, it is asked to take the proper subset $U$ of the natural join $U^*$, as a universal category relation, obtained by removing from $U^*$ the part (tuple) that can be inconsistent with the semantic constrains known to the statistical and medical analysts.

Notice that a tuple in **U** may refer to an empty category (accidentally empty category); such a tuple is called a dummy tuple. This distinction between structurally empty categories and accidentally empty categories which often occurs in the analysis of statistical data has to be reminded.

*Definition 2.3.* A universal table for a polyptych of summary tables is a summary table with classification scheme **U** and category relation $U$ such that its marginal over the $\mathbf{C}_i$'s coincide with the summary tables of the polyptych

*Definition 2.4.* A polyptych is consistent if it admits a universal table.

Due to assumptions 1 and 2, we eventually represent, in any Bioassay the universal tables as solutions of a linear constraint system.

A typical case in which a polyptych of summary table turns out to be consistent occurs when these come from a single data source. When summary tables are taken from distinct data sources, it is very improbable that the requirement of consistency will be fulfilled-inconsistency.

These points might be not so widely considered by an experimentalist, but the statistical techniques are strength with such considerations.

## 3. The covariates as extra categories

In most bioassays and at the experimental carcinogenesis as well, the target is to compare two different therapies/factors, so according to Section 1 it is $C = \{C_1, C_2\}$, or to evaluate the prognostic factors. But in principle, the population under study $\Omega$, is rather heterogeneous with respect to prognosis, it is asked to adjust the covariate effect describing the above mentioned heterogeneity, Cox and Snell (1989). Let $x_1$ be the factor of interest and $x_2$ the covariate and $\beta_1$, $\beta_2$ be the corresponding regression parameters. Fitting the full model with link function Link, McCullagh and Nedler (1989),

$$E(Y | x_1, x_2) = \text{Link}^{-1}(\beta_0 + \beta_1 x_1 + \beta_2 x_2), \tag{3.1}$$

while the restricted model with estimate $\beta_1^*$ is

$$E(Y | x_1) = \text{Link}^{-1}(\beta_0^* + \beta_1^* x_1). \tag{3.2}$$

Notice that the models (3.1) and (3.2) although nonlinear, are intrinsic linear. The corresponding variances to models (3.1) and (3.2) are:

$$\text{var}(Y | x_1, x_2) = \sigma_{Y.12}^2, \quad \text{var}(Y | x_1) = \sigma_{Y.1}^2 \tag{3.3}$$

Therefore the (relative) efficiency of $\hat{\beta}_1$ to $\hat{\beta}_1^*$ can be defined as

$$\text{eff}\left(\hat{\beta}_1, \hat{\beta}_1^*\right) = \frac{\sigma_{Y.12}^2}{\sigma_{Y.1}^2} = \frac{1 - \rho_{12}^2}{1 - \rho_{Y2.1}^2}, \tag{3.4}$$

with $\rho_{12} = \text{Corr}(x_1, x_2)$ and $\rho_{Y2.1}$ is the effect of $x_2$ on $Y$. From (3.4) is easy to see that

$$\text{eff}\left(\hat{\beta}_1, \hat{\beta}_1^*\right) = \begin{cases} 1, & \text{if } |\rho_{Y2.1}| = |\rho_{12}| \\ <1, & \text{if } \rho_{12} > \rho_{Y2.1} \\ >1, & \text{if } \rho_{12} < \rho_{Y2.1} \end{cases}. \tag{3.5}$$

In principle, interest is concentrated on a randomized treatment effect, i.e. $\rho_{12}=0$. That is emphasis is given in adjustment as eventually eff($\hat{\beta}_1, \hat{\beta}_1^*$)$\geq 1$. The question if $\beta_2 = 0$, which is actually a statistical null hypothesis, versus $\beta_2 \neq 0$ is crucial on misspecification by omitting or including $x_2$. Indeed: if $x_2$ is adjusted for, the assumed correct model (3.1) is fitted provided $\beta_2 \neq 0$. If $\beta_2 = 0$ this leads to overspecification of the model. If $x_2$ is not included, the (3.2) model is correct if $\beta_2 = 0$ while if $\beta_2 \neq 0$ the model (3.2) is underspecified.

If the link function is the logistic function, which remains invariant under certain transformation, Kitsos (2007a), the models (3.1) and (3.2) are reduced to

$$\log \frac{P_{1.12}}{1-P_{1.12}} = \beta_0 + \beta_1 x_1 + \beta_2 x_2, \tag{3.6}$$

$$\log \frac{P_{1.1}}{1-P_{1.1}} = \beta_0^* + \beta_1^* x_1, \tag{3.7}$$

with $P_{1,2} = P[Y=1|x_1,x_2]$ and $P_{1,1} = P[Y=1|x_1]$. In case that $\beta_1^* = \beta_1$ the plane (3.6) and the line (3.7) are parallel (Figure 1). The statistical implementation of this is equivalent that the RR for $x_1$ estimated either from (3.6) and (3.7) are equal as $RR = e^{\beta_1} = e^{\beta_1^*}$, otherwise the evaluated Relative Risk are non equal and is a matter of investigation. Notice that in both cases (3.6) and (3.7 are linear). If a second order model was consider, at least fir (3.7) the test for the curvature is really a necessity.

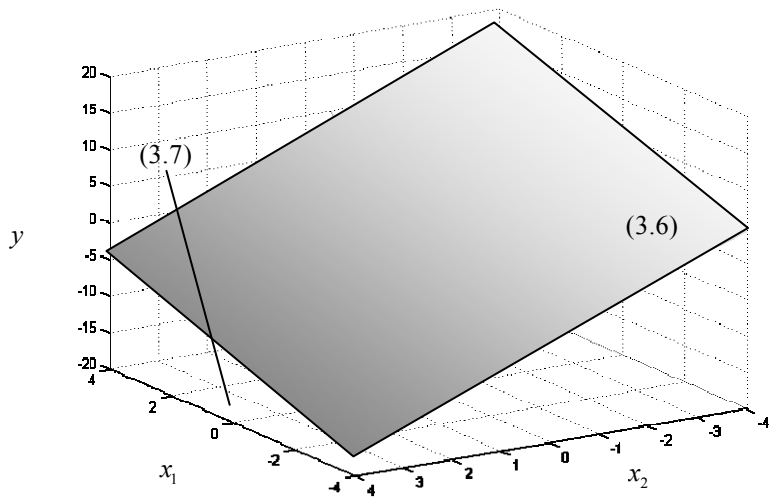

*Figure 1.* (3.6) *plane and* (3.7) *line.*

Notice that for the logistic model the curvature $1/Link(\cdot)$, is a convex leading function, to a downward bias of $\beta_1^*$ i.e. $\lim |\hat{\beta}_1^*| < |\beta_1|$, therefore the bias tends to zero only when $\beta_1 = 0$. Relative Risk to be equal to unit, i.e. $RR = 1$.

There is a similarity of the logistic and the Cox model, Prentice and Kalbfleisch (1979), Schumacher et al. (1987), Legakos and Schoenfeld (1984), while the behavior of the variances of the adjusted and unadjusted estimators need, we think, more investigation in this particular problem. We emphasize that the $\text{var}(\beta_0^*)$ and $\text{var}(\beta_1^*)$ in (3.7) can be reduced if a D-optimal or $D_s$-optimal design approach is adopted, Ford et al (1989), Kitsos (2007a).

## *4. Low-Dose models*

There are different statistical models to describe a process by which a normal cell becomes malignant through, at least one, transformation. When the malignancy is referred to a tumor we are referred to cancer. Then, the interest is focused on the "growth rate" of affected tissues as malignant tumors are capable of floating away and forming new malignant growths in other sites. Humans are certainly exposure to carcinogens, which is accepted for two main reasons:

- there are no reliable estimates for "safe doses" and
- the epidemiological methods are insensitive to small increase in cancer.

In quantitative toxicology the following definitions are adopted.

*Definition 4.1.* By "dose" we define the amount of chemical or energy in a radiological situation administered to or reviewed by exposed subject.

*Definition 4.2.* By "effect" we define an action as a result of a stimulus received through a receptor.

*Definition 4.3.* By "response" we mean any detectable change (and is assumed approached through a statistical model).

A number of extra definitions on the introduced term "dose" are considered. We briefly referred to them.

A "safe dose" it is assumed that will not increase the current cancer incidence rate by more than an "acceptable" low risk level, see the early work the Hartley and Sielken (1977). The "virtually safe dose" (VSD) was based on confidence intervals and it was rather mechanistic (Grump et al. (1977), Armitage (1982)). The upper confidence limit for the proportion of tumours was calculated and the dose-response curve was extrapolated towards zero.

Certainly one of the problems in the cancer risk assessment is the extrapolation from the experimental results to human Zapponi et al (1989). There are also different scale parameters and are discussed in comparisons of LD10≡$L_{-1}$, Brown and Howel (1983). These parameters are based on different species and a given set of chemicals. That is still interest is focused on low dose, which is equivalent to calculate the percentile $L_p$, $p \in (0,1)$ for "small" $p$, $0 < p \leq 0.1$, Kitsos (1997).

Although different terminology is adopted for the percentile point, like MTD, maximum tolerance dose, TD, tumorigenic dose, ED, estimated dose, LD, lethal dose the point remains the same: adopt that model which will provide the best downwards extrapolation to $L_p$, as it has no meaning to perform experiments below an unknown level of dose.

The crucial issue when fitting a model is prediction. The effect of covariates has to be considered, through the, assumed correct, model. The low-dose effects in a risk assessment have also to be within the study.

Different nonlinear models have been developed and applied under the name multistage models (MM). The class of MM has been mainly applied for the analysis of a large number of epidemiological data, Armitage (1985).The crucial issue when fitting a model from MM class is prediction. That is extrapolation downwards in the neighbourhood of zero, while prediction in statistics is rather related with a forward extrapolation. This is an interesting exception in the statistical General Model theory, where prediction is relating, mainly, in a forward extrapolation and in a lesser extend to the within the domain values.

There is a rather empirical approach to these models by toxicologists and environmentalists, without a reference to an explicit dose-response model, as tumour incidence data is titled to a prescribed dose-response relationship. These experimental approaches are still valuable and might provie interest results, Angelopoulou et al. (2008a, 2008b).

The dose-response curve, $F(\cdot)$ say, is a result of a binary response problem, recall Section 1, i.e. there are only two categories $C_1$ and $C_2$, see Kitsos (1998) for details. From the statistical point of view $F(x)$ is the cumulative distribution function, for the underlying probability model, describing the phenomenon. The notation x is referred to the dose level. However, due to a number of factors, including the temporal variability of animal population characteristics and the difficulty in identifying a specific animal breed, this dose-response curve is rather problematic to biologists and toxicologists. Moreover, F(x) is rather an assumed approximation, than a known deterministic mechanism for the phenomenon which describes. Therefore it needs an estimation and we would strongly recommend a Kolmogorov-Smirnov test.

When it is assumed that cancer is the result of a single event (or "hit") in a single cell, the one parameter exponential model

$$F(x) = 1 - \exp(-\theta x), \quad \theta > 0, \tag{4.1}$$

it is, known as *one-hit model*.

When a fixed number, say *k*, of (identical) "hits" occur in a tissue *the multi-hit model* is assumed to describe the phenomenon and the corresponding $F(x)$ is approximated by the assumed correct model

$$F(x) = \frac{1}{(k-1)!} \int_0^x u^{k-1} e^{-u} du. \tag{4.2}$$

When a sup linear relationship is assumed the one hit model is transformed to Weibull model with shape parameter, *s* say

$$F(x) = 1 - e^{-(\theta x)^s}.$$

The maximum likelihood estimation (MLE) for the parameters $(s, \theta)$, both assumed unknown, of the Weibull distribution is $L = L(s, \theta)$ and the log-likelihood $l = l(s, \theta)$.

Recall that the Weibull does not belong to the exponential family of models. The first derivatives are needed to evaluate the corresponding score functions

$$U_\theta = \frac{\partial l}{\partial \theta} = \frac{sd}{\theta} - s\theta^{s-1} \sum x_i^s, \tag{4.3}$$

$$\theta^* = \left( \frac{x}{\sum t_i^s} \right)^{1/s}. \tag{4.4}$$

When *s* is given, the MLE $\theta^*$ of $\theta$ can be found explicitly by solving $U_\theta = 0$ as

The second derivatives of the log-likelihood *l* are

$$I_{\theta\theta} = \frac{\partial^2 l}{\partial \theta^2} = \frac{\partial}{\partial \theta}\left( \frac{sd}{\theta} - s\theta^{s-1} \sum t_i^s \right) = -\frac{sd}{\theta^2} - s(s-1)\theta^{s-2} \sum t_i^s,$$

$$I_{\theta s} = \frac{\partial^2 l}{\partial \theta \partial s} = \frac{d}{\theta} - \theta^{s-1}(1 + s\log\theta) \sum t_i^s - s\theta^{s-1} \sum t_i^s \log t_i,$$

$$I_{ss} = \frac{\partial^2 l}{\partial s^2} = -\frac{d}{s^2} - \theta^s \sum t_i^s \left[ \log(\theta t_i) \right]^2.$$

Therefore Fisher's information matrix can be evaluated, as $I = I_{ij}$, $i,j = 1,2$ with

$$I_{11} = I_{\theta\theta}, \ I_{22} = I_{ss}, \ I_{12} = I_{21} = I_{\theta s} \ .$$

When it is assumed that the susceptible cell can be transformed through *k* distinct stages in order to be a malignant one the multistage model of Armitage-Doll (1954) described the phenomenon. The main assumption was that the transformation rate from each stage to the next on is linear. Eventually the cdf of developing cancer from exposure to a dose x, within a fixed time period, is given by

$$F(x) = 1 - \exp[-(\theta_0 + \theta_1 x + ... + \vartheta_k x^k)] , \qquad (4.5)$$

where $\theta_i$, $i = 0,1,...,k$ are defined through the coefficients of the linear transformations assumed between stages, Grumb et al. (1977), i.e. $\vartheta_i = \vartheta_i(t)$. The most usual model are the multistage linear model and the multistage model. Notice that model (4.5) developed on a completely different biological insight and not as general mathematical form of the previous models.

The Logit and Probit models, Mc Cullagh and Nedler (1993), known as tolerance distribution models in cancer risk assessments, are also useful to toxicology and are included to MM class.

In pharmacokinetics for cancer risk assessment the Michaelis-Menten metabolic process is usually considered when it is assumed to lead to a concentration of the active metabolite in the target tissue considering as function of x, see Appendix 1 for more details.

The MM class is the earlier appeared, Armitage and Doll (1954), and is based on the assumption that a single normal cell may become fully malignant when a sequence of say *k*, irreversible heritable mutation-like changes assumed. Now, under the assumption that the intermediate cells are subject to a stochastic birth-death process for cell proliferation and cell differentiation, when $k = 2$ the Biologically Based Models (BBM) was created by Moolgavkar and his associates, see Moolgavkar and Venson (1979) and developed by a series of papers by Luebeck and Moolgavkar (1989, 1991, 1992).

The two families of models, the MM and BBM, are based on different hazard functions. Indeed if the mutation rates are very small and independent of time the hazard function of cancer for the Armitage–Doll model is

$$\lambda(t) = c(t - t_0)^{k-1}, \quad c > 0, \tag{4.6}$$

where $k$ is the number of stages, $t_0$ is a fixed and positive number for the growth of tumour.

When interest is focused to identify etiological agents of cancer and develop the appropriate statistics for risk assessment of environmental agents then the most appropriate hazard function is the one defined by Cox (1972) as

$$\lambda(t) = \lambda_0(t) S(W, \beta), \tag{4.7}$$

with $\lambda_0(t) > 0$ known as baseline hazard function, $S(W, \beta)$ the risk function which relates the environmental factor $W$, i.e. the covariates and the vector of unknown parameters $\beta$. This model is known as a proportional-hazard model.

An interesting application of the proportional-hazard models has been discussed by Pargament et al. (2001). He worked on religious struggle as a predictor of mortality among sick patients. His data set was based on 576 Baptists and Methodists, age over 55, hospitalized in a particular hospital, and they were follow-up for two years with 176 deaths and 152 subjects were lost to follow-up. Adopting o proportional-hazard model an interesting, rather social than medical analysis is presented. A well known technique for cancer is screening. If screening speeds-up detection that will eventually increase the time (known as "lead time") from detection to death. The lead time for the breast cancer screening was discussed by Patz et al. (2000), Welch at al. (2007).

Proportional-hazard models are mainly applied in clinical trials. In principle, in a clinical trial we need to know a curve for the treatment group and another one for the control group, due to Kaplan-Meier estimator. If the treatment is not depended on failure time the corresponding survival curve will fall off slowly, while if the treatment has no effect the two curves will statistically coincide. The above discussion only tries to encourage that the mathematical formulation, does not solve the problem, it describes it. An essential analysis is needed as the conclusions are rather sensitive, concerning human lives.

The MMB are based on a Poisson process for stage to stage. For example Moolgarkar and Venzon (1979) assumed a Poisson process with birth rate at *i* cell $b_i(t) = ib$ and death rate at *i* cell $d_i(t) = id$, i.e. a homogenous birth-rate process.

## *5. Nonlinear Models*

In this section we briefly discuss typical non-linear models which might provide response curves with no significant difference between them are, with the same parameter vector. But when the parameter vector is based on different values the same non-linear might appear close to a line. That is why we review these models, we evaluated their graphs in Appendix B were these graphs, within an MS Excel environment, can easily provide the curve by changing the initial guesses for the estimators. For a number of these models the Fisher's information matrix is evaluated as $i(\theta) = (\nabla f)^T (\nabla f) \sigma^{-2}$ for an observation. In other cases, the partial derivatives are evaluated so that to form $i(\theta)$, see Kitsos (2007b).

| MODEL | NAME |
|---|---|
| **1.** $f_G(u,\theta) = \theta_0 \exp(\theta_1 e^{\theta_2 u})$ | : Gompetz model. |
| **2.** $f_J(u,\theta) = \theta_0 + \theta_1 \exp(\theta_2 u^{\theta_3})$ | : Janoscheck model. |
| **3.** $f_L(u,\theta) = \theta_0 / (1 + \theta_1 e^{\theta_2 u})$ | : Logistic model. |
| **4.** $f_B(u,\theta) = [\theta_0 + \theta_1 e^{\theta_2 u}]^3$ | : Bertalanffy model. |
| **5.** $f_{\tanh}(u,\theta) = \theta_0 + \theta_1 \tanh(\theta_2(u - \theta_3))$ | : tahn-model. |
| **6.** $f_{3\text{-}\tanh}(u,\theta) = \dfrac{\theta_0}{2}\left[1 + \dfrac{2}{\pi}\arctan(\theta_1(u - \theta_2))\right]$ | : 3-tanh-model. |
| **7.** $f_{4\text{-}\tanh}(u,\theta) = \theta_0 + \dfrac{2}{\pi}\theta_1 \arctan(\theta_2(u - \theta_3))$ | : 4-tanh-model. |
| **8.** $f_{\exp}(u,\theta) = \theta_0 u^{\theta_1} = \vartheta_0 e^{\theta_1 \ln u}$ | : Exponential time-power model. |
| **9.** $f_{2\exp}(u,\theta) = \theta_0 - \theta_1 e^{-\theta_2 \ln u}$ | : Reparametrized Exponential time-power model. |
| **10.** $f_W(u,\theta) = \theta_0 - (\theta_0 - \theta_1)\exp(-(\theta_2 u)^{\theta_3})$ | : Reconstructed Weibull model. |
| **11.** $f_{GL}(u,\theta) = \theta_0 / \left[1 + e^{\theta_1 + \theta_2 g(u)}\right]$ | : Generalized Logistic model |

We evaluate for various models parameters Fisher's Information Matrix for one observation is $i(\theta) = (\nabla f)^T (\nabla f) \sigma^{-2}$. Indeed:

a. For the Gompetz model,

$$i(\theta) = \frac{u^{2\theta_1}}{\sigma^2} \begin{bmatrix} 1 & \theta_0 \ln u \\ \theta_0 \ln u & \theta_0^2 (\ln u)^2 \end{bmatrix}.$$

b. For the reparametrized Exponential time-power model,

$$i(\theta) = \sigma^{-2} \begin{bmatrix} 1 & -e^{\theta_2 u} & \theta_1 u e^{-\theta_2 u} \\ -e^{\theta_2 u} & e^{-2\theta_2 u} & -\theta_1 u e^{-2\theta_2 u} \\ \theta_1 u e^{-\theta_2 u} & -\theta_1 u e^{-2\theta_2 u} & (\theta_1 u)^2 e^{-2\theta_2 u} \end{bmatrix}.$$

c. For the reconstructed Weibull model, the partial derivatives are:

$$\frac{\partial f_W}{\partial \theta_0}(u, \theta) = 1 - e^{-(\theta_2 u)^{\theta_3}}, \quad \frac{\partial f_W}{\partial \theta_1}(u, \theta) = -e^{-(\theta_2 u)^{\theta_3}},$$

$$\frac{\partial f_W}{\partial \theta_2}(u, \theta) = (\theta_0 - \theta_1) e^{-(\theta_2 u)^{\theta_3}} u^{\theta_3} \theta_2^{\theta_3 - 1} \quad (\theta_2 > 0),$$

$$\frac{\partial f_W}{\partial \theta_3}(u, \theta) = (\theta_0 - \theta_1) e^{-(\theta_2 u)^{\theta_3}} (\theta_2 u)^{\theta_3} \ln(\theta_2 u) \quad (\theta_2 u > 0).$$

Therefore, $i(\theta) = \sigma^{-2} [i_{kl}]_{k,l \in \mathbb{N}_1^4}$, where

$$i_{11} = \left[1 - e^{-(\theta_2 u)^{\theta_3}}\right]^2, \quad i_{22} = e^{-2(\theta_2 u)^{\theta_3}},$$

$$i_{33} = \left[\frac{\theta_0 - \theta_1}{\theta_2} e^{-(\theta_2 u)^{\theta_3}} (\theta_2 u)^{\theta_3}\right]^2, \quad i_{44} = (\theta_0 - \theta_1)^2 (\ln \theta_2 u)^2 e^{-2(\theta_2 u)^{\theta_3}} (\theta_2 u)^{2\theta_3},$$

$$i_{12} = i_{21} = -\left[1 - e^{-(\theta_2 u)^{\theta_3}}\right] e^{-(\theta_2 u)^{\theta_3}},$$

$$i_{13} = i_{31} = \left[1 - e^{-(\theta_2 u)^{\theta_3}}\right] \frac{\theta_0 - \theta_1}{\theta_2} e^{-(\theta_2 u)^{\theta_3}} (\theta_2 u)^{\theta_3},$$

$$i_{14} = i_{41} = \left[1 - e^{-(\theta_2 u)^{\theta_3}}\right] (\theta_0 - \theta_1) \ln(\theta_2 u) e^{-(\theta_2 u)^{\theta_3}} (\theta_2 u)^{\theta_3},$$

$$i_{23} = i_{32} = -e^{-(\theta_2 u)^{\theta_3}} \frac{\theta_0 - \theta_1}{\theta_2} e^{-(\theta_2 u)^{\theta_3}} (\theta_2 u)^{\theta_3},$$

$$i_{24} = i_{42} = -e^{-(\theta_2 u)^{\theta_3}} (\theta_0 - \theta_1) \ln(\theta_2 u) e^{-(\theta_2 u)^{\theta_3}} (\theta_2 u)^{\theta_3},$$

$$i_{34} = i_{43} = \frac{(\theta_0 - \theta_1)^2}{\theta_2} \ln(\theta_2 u) e^{-2(\theta_2 u)^{\theta_3}} (\theta_2 u)^{2\theta_3}.$$

**d.** For the Generalized Logistic model, we have the following cases:

Case **(i)**: $g(u) = \theta_1 u + \theta_2 u^2 + \theta_3 u^3$, and therefore

$$f_{GL}(u,\theta) = \frac{\theta_0}{1 + e^{\theta_1 + \theta_2 u + \theta_3 u^2 + \theta_4 u^3}}.$$

For this case the partial derivatives are:

$$\frac{\partial f_{GL}}{\partial \theta_0}(u,\theta) = \frac{1}{1 + e^{\theta_1 + \theta_2 u + \theta_3 u^2 + \theta_4 u^3}},$$

$$\frac{\partial f_{GL}}{\partial \theta_1}(u,\theta) = -\frac{\theta_0 e^{\theta_1 + \theta_2 u + \theta_3 u^2 + \theta_4 u^3}}{\left(1 + e^{\theta_1 + \theta_2 u + \theta_3 u^2 + \theta_4 u^3}\right)^2},$$

$$\frac{\partial f_{GL}}{\partial \theta_2}(u,\theta) = -\frac{\theta_0 u e^{\theta_1 + \theta_2 u + \theta_3 u^2 + \theta_4 u^3}}{\left(1 + e^{\theta_1 + \theta_2 u + \theta_3 u^2 + \theta_4 u^3}\right)^2},$$

$$\frac{\partial f_{GL}}{\partial \theta_3}(u,\theta) = -\frac{\theta_0 u^2 e^{\theta_1 + \theta_2 u + \theta_3 u^2 + \theta_4 u^3}}{\left(1 + e^{\theta_1 + \theta_2 u + \theta_3 u^2 + \theta_4 u^3}\right)^2},$$

$$\frac{\partial f_{GL}}{\partial \theta_4}(u,\theta) = -\frac{\theta_0 u^3 e^{\theta_1 + \theta_2 u + \theta_3 u^2 + \theta_4 u^3}}{\left(1 + e^{\theta_1 + \theta_2 u + \theta_3 u^2 + \theta_4 u^3}\right)^2}.$$

Case **(ii)**: $g(u) = \dfrac{u^{\theta_3} - 1}{\theta_3}$, and therefore

$$f_{GL}(u,\theta) = \frac{\theta_0}{1 + e^{\theta_1 + \theta_2 \frac{u^{\theta_3} - 1}{\theta_3}}}.$$

For this case the partial derivatives are:

$$\frac{\partial f_{GL}}{\partial \theta_0}(u,\theta) = \frac{1}{1+e^{\theta_1+\theta_2\frac{u^{\theta_3}-1}{\theta_3}}},$$

$$\frac{\partial f_{GL}}{\partial \theta_1}(u,\theta) = -\frac{\theta_0 e^{\theta_1+\theta_2\frac{u^{\theta_3}-1}{\theta_3}}}{\left(1+e^{\theta_1+\theta_2\frac{u^{\theta_3}-1}{\theta_3}}\right)^2},$$

$$\frac{\partial f_{GL}}{\partial \theta_2}(u,\theta) = -\frac{\theta_0 (u^{\theta_3}-1)e^{\theta_1+\theta_2\frac{u^{\theta_3}-1}{\theta_3}}}{\theta_3\left(1+e^{\theta_1+\theta_2\frac{u^{\theta_3}-1}{\theta_3}}\right)^2},$$

$$\frac{\partial f_{GL}}{\partial \theta_3}(u,\theta) = -\frac{\theta_0 (u^{\theta_3}-1)\theta_2 u^{\theta_3-1}\left[\theta_3(\ln u)-1\right]e^{\theta_1+\theta_2\frac{u^{\theta_3}-1}{\theta_3}}}{\theta_3^2\left(1+e^{\theta_1+\theta_2\frac{u^{\theta_3}-1}{\theta_3}}\right)^2}.$$

Then we can evaluate the Fisher's Information matrix $i(\theta)$. The information matrix of this model does not depend on the linear added term. That is, in principle, for the model $g(u,\theta)$ and $f(u,\theta) = \theta_0 + g(u,\theta)$ the estimated Fisher's information matrix needs prior information of the parameters involved in $g(u,\theta)$, not for $\theta_0$. The estimation of $\sigma^2$, $s^2$ is also needed so that to have an estimate $\hat{i}(\hat{\theta}) = (\nabla f)^T (\nabla f) s^{-2}\mid_{\theta=\hat{\theta}}$.

## *Conclusions*

There is a theoretical background to cover the performance of any Bioassay, so for a cancer one. Not only to impose the appropriate formulation to group the data set in Tables as far as descriptive statistics concern. Interest was also on how to use the appropriate non-linear usually model. This model can be either a binary response one, or any other non-linear model in the continuous case. We provided a critical view of this analysis and, we believe, we offer this appropriate background to experimentalists. So the link between statistical/mathematical model and a cancer bioassay to be better bridged.

## *Appendix A*

- *More about Michaelis-Menten model*

The biochemical model for a simple enzyme-substate reaction, derivate by Michaelis-Menten has various form extension.

**I.** Consider the reaction scheme

$$E + S \underset{k_2}{\overset{k_1}{\rightleftharpoons}} ES \overset{k_3}{\longrightarrow} E + P, \quad \text{(A1.1)}$$

with $E$: enzyme, $S$: its substrate, $P$: product of the reaction, $k_1, k_2, k_3$: rate constants. In the steady state concentration, denoted by [], of $ES$ is constant so that

$$k_1[E][S] = (k_2 + k_3)[ES]. \quad \text{(A1.2)}$$

If $E_0$ is the total concentration of enzyme present, actually independent of time, then

$$E_0 = [E] + [ES].\tag{A1.3}$$

So (2) from (3) becomes

$$k_1[s]E_0 = (k_1[S] + k_2 + k_3)[ES] \Rightarrow [ES] = \frac{k_1[S]E_0}{k_1[S] + k_2 + k_3}.\tag{A1.4}$$

Equation (4) presents a rectangular hyperbola in $[S]$, and provides the concentration of substrate molecules that are combined with enzyme molecules. The older than MM known as Langmuir´s absorption isotherm was of the same form. It was referring to the absorption of gas molecules on the solid surfaces.

Let as demote by $v$ the speed of the steady-state reaction $v = k_3[ES]$. Then from (4) we get

$$v = \frac{k_1 k_3 E_0 [S]}{k_1[S] + k_2 + k_3} := \frac{V_{max}[S]}{K + [S]}.\tag{A1.5}$$

With

$$V_{max} = k_3 E_0, \quad K = \frac{k_2 + k_3}{k_1} := K_M := K_{max},\tag{A1.5a}$$

the MM constant $K_M \equiv K \equiv K_{max}$ is the value of substrate concentration for the half-maximal velocity, $V = V_{max}/2$.

The $V_{max}$ denote the maximum velocity of the reaction and is obtained when all the active sites on the enzyme molecules are occupied by substrate molecules.

From (5) it is easy to see that the passes through the origin when $[S] = 0$. The slope is

$$\left.\frac{dv}{d[S]}\right|_{[S]=0} = \frac{V_{max}}{K}.$$

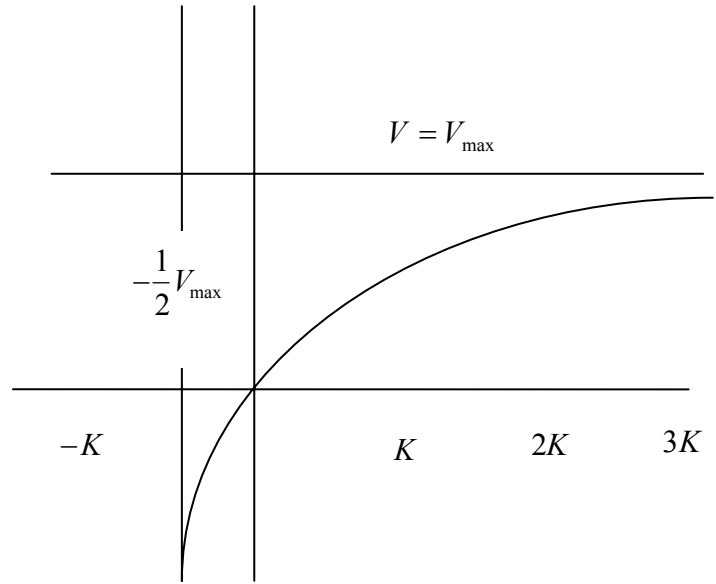

*Figure A.1. The Michelis-Menten model.*

Notice that, if we let $x = [S]$,

$$\frac{dv}{dx} = \frac{V_{max} K_M}{(K_M + x)^2}, \quad \frac{d^2 v}{dx^2} = -\frac{2 V_{max} K_M}{(K_M + x)^3}.$$  (A1.7)

The rectangular hyperbola or MM needs two parameters to be specified: those which define the asymptotes or equivalently the parameters of the model.

- Applications of Rectangular Hyperbola (single substrate)
- The specific growth rates of micro-organisms [Dean-Hinshelwood, pg. 80-81]
- The respiration rate of mature leaves [Yemm, pg. 275]
- The respiration in potatoes [Barker]

It is interest to notice that leaf photosynthesis, among other physiological problems, are considered as non-rectangular hyperbolic responses with single substrate, while rectangular hyperbolas for two substrates of the form

$$v = \frac{k x_1 x_2}{1 + C_1 x_1 + C_2 x_2 + C_3 x_1 x_2}.$$  (A1.8)

The equation (5) describes two-substrate enzyme kinetics under certain conditions; see Dixon-Webb, pg. 100.

So (8) is a more general form of (5) or any rectangular hyperbola of the form

$$\frac{ax}{b+x}, \text{ with } a, b \text{ constants.}$$

**II.** Now, modify the MM in the form

$$E + 2X \rightleftharpoons EX_2 \longrightarrow E + product,$$

where $E$ : enzyme, $X$ : substrate.

In such a case molecules of $X$ can combine with $E$ only two at the time. It can be shown that the utilization rate $V$ of substrate $X$ is given by

$$U = \frac{Vx^2}{K^2 + X^2}. \tag{A1.9}$$

Sometimes a factor $X$ is above a critical value $K_c$.

The response can be derived from molecular models and it can be of the form

$$U = \frac{Vx^n}{K_c^n + X^n}, \tag{A1.10}$$

with $K_c$ the threshold, $U$ the response, $X$ the density concentration level of some substance $K_c$ : value of $X$ for half-maximal response, $n$ : usually positive integer and $V$ a constant.

$$\frac{U}{V} = \frac{(X/K_c)^n}{1 + (X/K_c)^n}. \tag{A1.11}$$

Another model can be

$$\frac{U}{V} = \frac{1}{1 + (X/K_c)^n}. \tag{A1.11a}$$

Model (A1.11) is of the same form of the Morgan-Marcer-Flodin (MMF) family of models

$$f_{MMF}(u, \theta) = \frac{\theta_0 x^{\theta_1} + \theta_2 \theta_3}{x^{\theta_1} + \theta_3}, \tag{A1.11b}$$

see Seber and Wild (1989, pg. 342) for details.

- *MM process in PARALLEL*

The overall behavior of two MM transport process working independently in parallel process. The total flux density M is assumed to be

$$MM = CMM_1 + CMM_2 = \frac{V_1[S]}{K_1+[S]} + \frac{V_2[S]}{K_2+[S]},$$

with $CMM_1$ contributed $MM_1$, $CMM_2$ contributed $MM_2$, $V_1, V_2, K_1, K_2$ constants.

$$\frac{d}{d[S]} M = \frac{V_1 K_1}{(K_1+[S])^2} + \frac{V_2 K_2}{(K_2+[S])^2},$$

$$\frac{d^2}{d[S]^2} M = -\frac{2V_1 K_1}{(K_1+[S])^3} - \frac{2V_2 K_2}{(K_2+[S])^3},$$

$$\frac{d}{d[S]} M \bigg|_{[S]=0} = \frac{V_1}{K_1} + \frac{V_2}{K_2} \qquad M \bigg|_{[S] \to \infty} \approx V_1 + V_2.$$

- *MM processes in SERIES*

$$E + S \underset{k_2}{\overset{k_1}{\rightleftharpoons}} ES \underset{k_4}{\overset{k_3}{\rightleftharpoons}} E + I, \qquad (A1.13)$$

where $E, S, I$ are enzymes,

$$MM = \frac{E_0(k_1 k_3[S] - k_1 k_3[I])}{k_2 + k_3 + k_1[S] + k_4[I]},$$

where $E_0$: total concentration of enzyme present.

- *Photosynthetic Response Light and $CO_2$*

$$P_{max} = \frac{P_0 \eta C}{P_0 + \eta C} = \frac{P_0 u}{P_0 + u}, u = \eta C.$$

- *Leaf Response to light flux density $I_t$:*

$$P_\eta = \frac{a I_t P_{max}}{a I_t + P_{max}} - R_d.$$

## *Appendix B*

The figures presented here are evaluated with parameters equal to 1. However, in any of the following models, the reader can examine how a model behaves by changing the appropriate $\theta_i$'s and then can observe corresponding change in the related figure (in the MS Word document of this paper double-click on $\theta_i$'s in order to change them, and then right-click on the related figure below and choose "update link" in order to update the figure. Thus, the reader needs to have the MS Word document of this paper and the additional Graph.xls file –provided by the authors upon request– in the same folder).

- $f_G(u,\theta) = \theta_0 \exp(\theta_1 e^{\theta_2 u})$ .

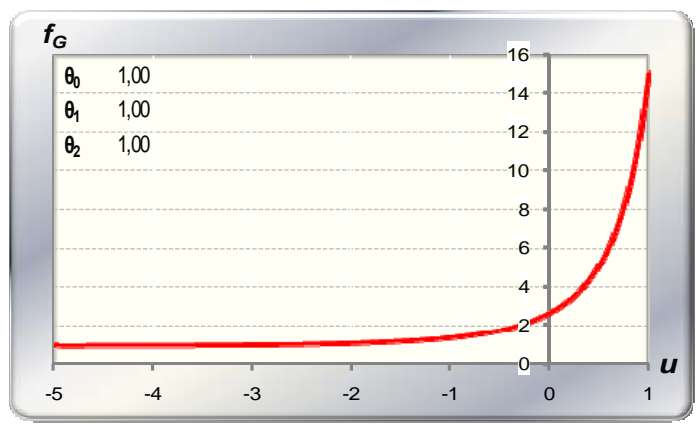

*Figure B.1. Gompetz model.*

- $f_J(u,\theta) = \theta_0 + \theta_1 \exp(\theta_2 u^{\theta_3})$ .

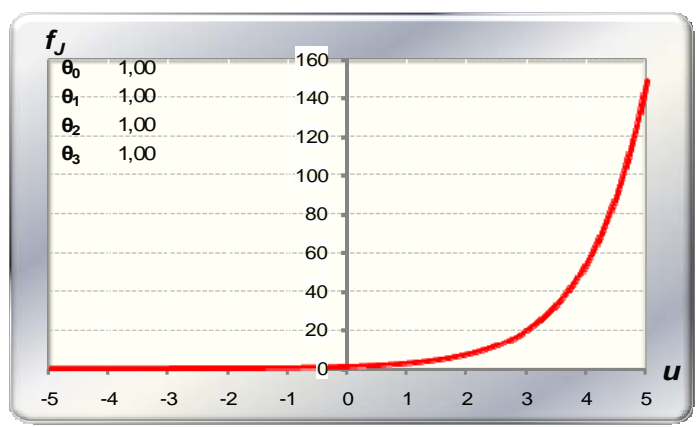

*Figure B.2. Janoscheck model.*

- $f_L(u,\theta) = \theta_0/[1+\theta_1\exp(\theta_2 u)]$.

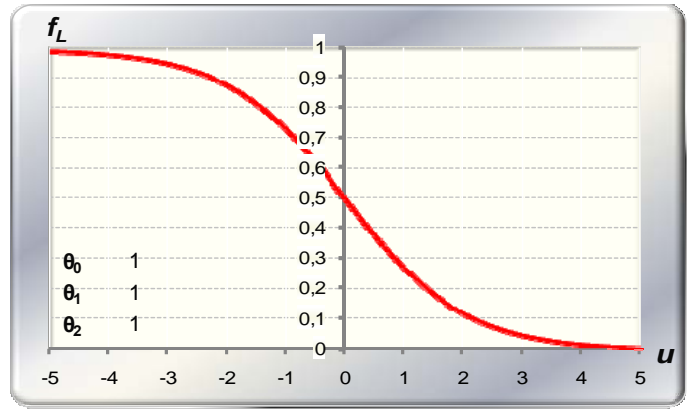

*Figure B.3. Logistic model.*

- $f_B(u,\theta) = [\theta_0 + \theta_1\exp(\theta_2 u)]^3$.

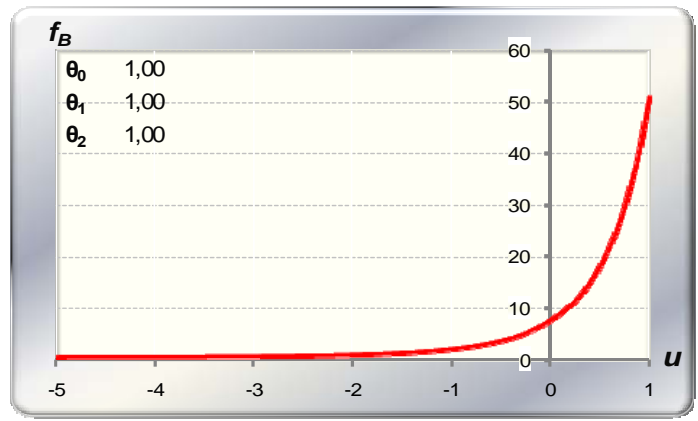

*Figure B.4. Bertalanffy model.*

- $f_J(u,\theta) = \theta_0 + \theta_1 \exp(\theta_2 u^{\theta_3})$, $f_B(u,\theta) = [\theta_0 + \theta_1 \exp(\theta_2 u)]^3$.

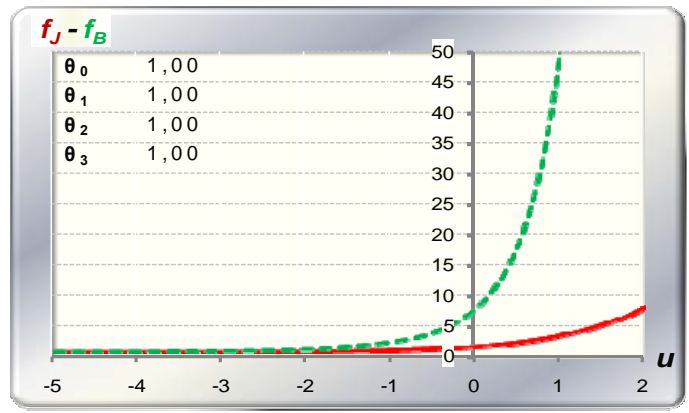

*Figure B.5. Comparison of Janoscheck and Bertalanffy models.*

- $f_{\tanh}(u,\theta) = \theta_0 + \theta_1 \tanh(\theta_2(u - \theta_3))$.

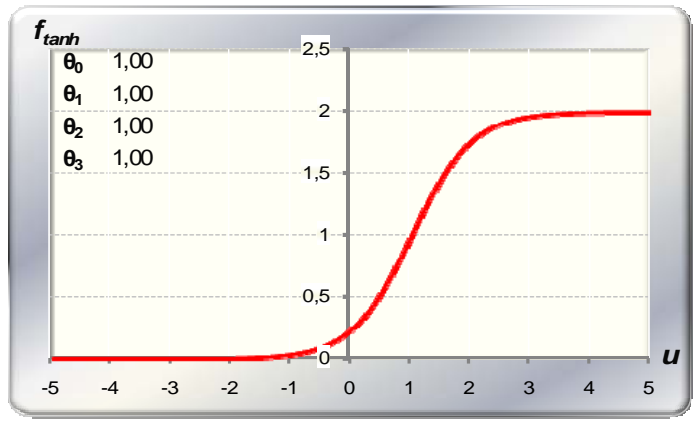

*Figure B.6. tanh-model.*

- $f_{3\text{-tanh}}(u,\theta) = \dfrac{\theta_0}{2}\left[1 + \dfrac{2}{\pi}\arctan\left(\theta_1(u-\theta_2)\right)\right].$

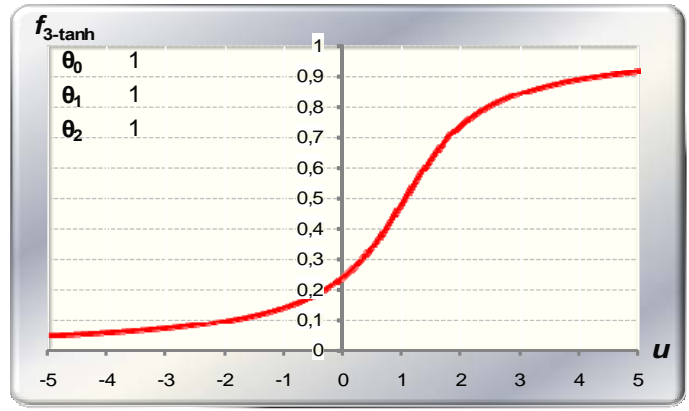

*Figure B.7. 3-tanh-model.*

- $f_{4\text{-tanh}}(u,\theta) = \theta_0 + \dfrac{2}{\pi}\theta_1 \arctan\left(\theta_2(u-\theta_3)\right).$

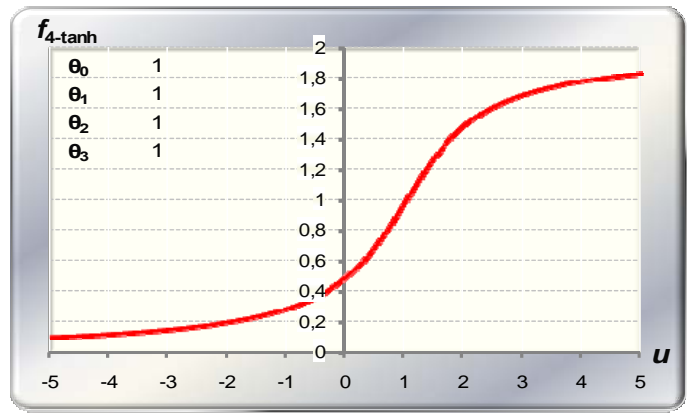

*Figure B.8. 4-tanh-model.*

- $f_{\tanh}(u,\theta) = \theta_0 + \theta_1 \tanh(\theta_2(u - \theta_3))$,

  $f_{3\text{-}\tanh}(u,\theta) = \dfrac{\theta_0}{2}\left[1 + \dfrac{2}{\pi}\arctan(\theta_1(u - \theta_2))\right]$,

  $f_{4\text{-}\tanh}(u,\theta) = \theta_0 + \dfrac{2}{\pi}\theta_1 \arctan(\theta_2(u - \theta_3))$.

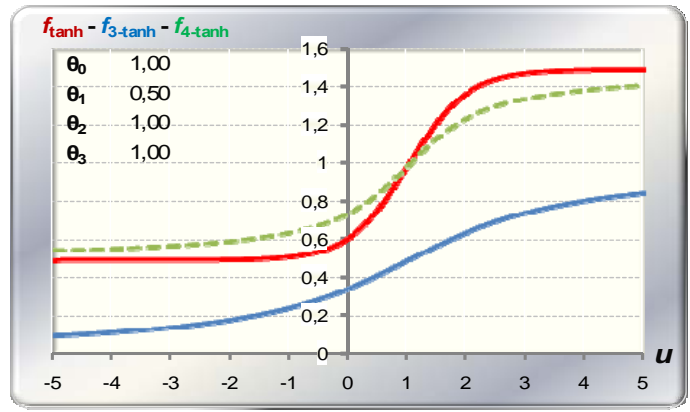

*Figure B.9. Comparison of tanh-, 3-tanh- and 4-tanh-models.*

- $f_{\exp}(u,\theta) = \theta_0 u^{\theta_1} = \vartheta_0 e^{\theta_1 \ln u}$, $f_{2\exp}(u,\theta) = \theta_0 - \theta_1 e^{-\theta_2 \ln u}$.

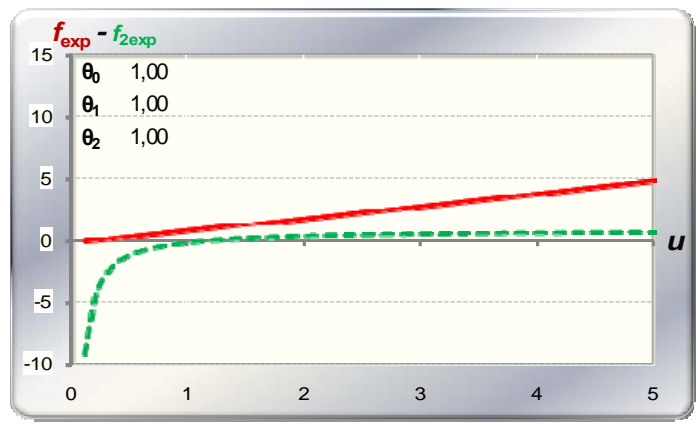

*Figure B.10. Comparison of exponential and reparametrized exponential time-power models.*

- $f_W(u,\theta) = \theta_0 - (\theta_0 - \theta_1)\exp\left(-(\theta_2 u)^{\theta_3}\right)$.

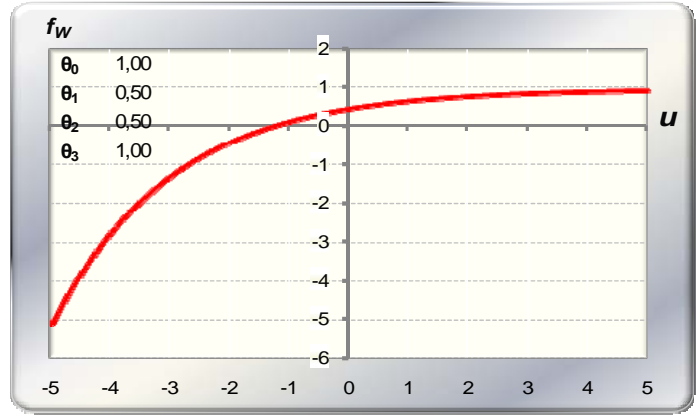

*Figure B.11. Reconstructed Weibull model.*

- $f_{GL1}(u,\theta) = \dfrac{\theta_0}{1+e^{\theta_1+\theta_2 u+\theta_3 u^2+\theta_4 u^3}}$, $f_{GL2}(u,\theta) = \dfrac{\theta_0}{1+e^{\theta_1+\theta_2 \frac{u^{\theta_3}-1}{\theta_3}}}$.

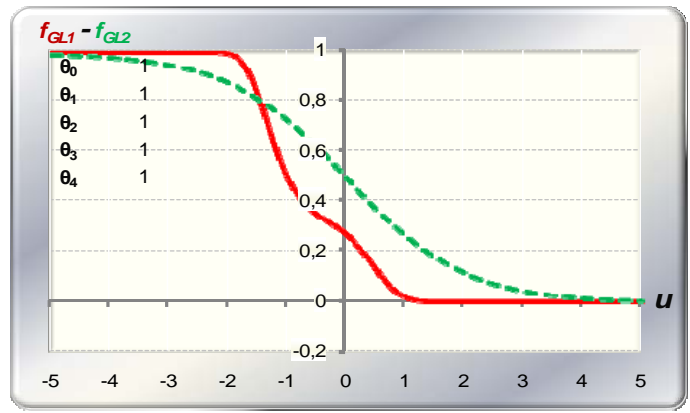

*Figure B.12. Generalized Logistic models (cases i and ii).*